\documentclass[a4paper, amsfonts, amssymb, amsmath, reprint, showkeys, nofootinbib, twoside]{revtex4-1}
\usepackage[english]{babel}
\usepackage[utf8]{inputenc}
\usepackage[colorinlistoftodos, color=green!40, prependcaption]{todonotes}
\begin{document}

\title{Generalized Abraham Forces on Molecules: from classical to quantum}
\author{B.A. van Tiggelen}
    \email[Correspondence email address: ]{bart.van-tiggelen@grenoble.cnrs.fr}
    \affiliation{Univ. Grenoble Alpes, CNRS, LPMMC, 38000 Grenoble, France}

\date{\today} 

\begin{abstract}
 Abraham forces are defined as electromagnetic forces on neutral objects caused by the presence of slow, time-dependent, homogeneous electromagnetic fields.
 Only a few experimental observations have been reported, and different formulations exist in literature. The ``standard" Abraham force is usually associated with the full derivative $\partial_t(\mathbf{E} \times \mathbf{B})$ of homogeneous electric and magnetic fields. We show that this full derivative is wrong in general, since the classical Maxwell theory contains subtle induced-dipole forces that contribute to the Abraham force, which thus breaks the dual symmetry between electric and magnetic components. We investigate other manifestations of Abraham forces and torques and estimate orders of magnitude, e.g related to chiral, moving or rotating matter. Some of them are also severely violating dual symmetry. We  provide a quantum description of the Abraham force. Apart from being more elegant and less technical, this quantum treatment paves the way for future studies of Abraham forces on complex molecules involving spin and spin-orbit coupling, and also facilitates the inclusion of the quantum vacuum.
\end{abstract}


\maketitle

\section{Introduction}

Electromagnetic fields exert a force on charges and moving charges. This is the celebrated Lorentz force, and understood, both classically, quantum-mechanically and relativistically. The force on small, electrically neutral particles though with permanent electric ($\mathbf{d}$) or magnetic ($\mathbf{m}$) dipole moment is  found in text books to be $d_i\nabla E_i$  and $m_i\nabla B_i$, respectively \cite{Jackson}. The same is true for the torque exerted by the electromagnetic field, equal to $\mathbf{d}\times \mathbf{E}$ and $\mathbf{m}\times \mathbf{B}$. More controversial however, are the force and torque exerted on particles with no net charge, current or dipole moment. Because the dipole moments now have to be induced, this force is at least quadratic in the electromagnetic fields, involving in principle either their spatial gradient or time-derivative. The Abraham force is often defined as the force exerted by homogeneous, slow, time-dependent fields on neutral, polarizable and/or magnetizable particles. If we focuss on the combinations  of time-derivatives of  homogeneous electric and magnetic fields alone, we can propose an ``Abraham force" of the type
 \begin{equation}\label{Fa}
    \mathbf{F}_A^{(1)} = \gamma_E \partial_t \mathbf{E}_0 \times \mathbf{B}_0 + \gamma_B \mathbf{E}_0 \times \partial_t \mathbf{B}_0
\end{equation}
featuring two material parameters $\gamma_{E,B}$, with apparently no further symmetry constraints. Except for the version by Peierls \cite{Peierls}, most work agrees on $\gamma_E=\gamma_B$, i.e. implying a full derivative $\partial_t(\mathbf{E}_0 \times \mathbf{B}_0)$, but disagree on the exact value \cite{Brevik}. A full time-derivative $\mathbf{F}_A = d\mathbf{p}_A/dt$ would imply the identification of a  momentum $\mathbf{p}_A$ that contributes to the total momentum of matter and radiation. A longstanding Abraham-Minkowski controversy exists on what is the exact momentum of a photon in matter as summarized by the thorough review by Brevik \cite{Brevik} that also emphasizes that this question closely relates to the  Abraham force in Eq.~(\ref{Fa}).  Several solutions have been offered \cite{AbMiBarnett,MinusipurAbMi,Nelson}. An interesting, additional dual symmetry between electric and magnetic fields has been proposed \cite{dual,dual2} that leads to the version due to Einstein and Laub in which momentum density energy and current density coincide op to a factor $c_0^2$ \cite{barnettloudon}. Dual symmetry is manifestly obeyed by the Maxwell equations in vacuum, but when electromagnetic waves interact with matter it may be broken. As in the Minkowski version, the Einstein/Laub version predicts $\gamma_E=\gamma_B \sim \varepsilon - 1/\mu$ in matter with dielectric constant $\varepsilon$ and magnetic permeability $\mu$.

Abraham forces are very small but touch the heart of electromagnetism, because of their relation with momentum and angular momentum of electromagnetic waves, a relation that is clearly more explicit in the quantum formulation. For instance, Barnett etal. \cite{AbMiBarnett} demonstrate that the difference between the Abraham and the Minkowski momentum coincides with the difference between kinetic momentum and canonical momentum in the electric-dipole gauge, the latter being associated with the De Broglie wavelength. The first observation of the Abraham force goes back to Walker etal.\cite{Walker} who measured the Abraham torque $\mathbf{r} \times  (\gamma_E \partial_t \mathbf{E}_0 \times \mathbf{B}_0)$ exerted on a strong ceramic capacitor using  electric fields oscillating at frequencies as low as  $0.4$ Hz, yielding forces of order $F_A = 20 $ nN. Rikken and Van Tiggelen \cite{Geert1} reported Abraham forces on molecules in gaze phase of order $F_A = 160 $ nN/mole with electric fields at 8 kHz and constant magnetic fields and found perfect proportionality to the static electric polarizability. Since no magnetic permeability exist, this observation is consistent with both the Abraham and Einstein/Laub versions, that favor $\mathbf{E}\times\mathbf{ H}/c_0^2$ for the momentum density, and the Nelson version that advocates $\mathbf{E}\times\mathbf{B}/c_0^2$ \cite{Nelson}. It is known that the Abraham force is subject to contributions from the quantum vacuum \cite{Feigel,kawka}. Ref.~\cite{Geert1} tried to observe, in vain, the quite significant QED correction predicted by Feigel \cite{Feigel}. The first work that employed both oscillating electric and magnetic fields exposed on a ceramic with huge dielectric constant reported large Abraham forces of order $1\, \mu$N per Tesla at frequencies up to 1 kHz \cite{Geert2}, and found that $\gamma_E\approx \gamma_B$.

One can speculate about other kinds of Abraham forces featuring only electric or magnetic fields, allowed by $P$,$C$ and $T$-symmetry in Maxwell equations, such as
 \begin{equation}\label{Fb}
    \mathbf{F}_A^{(2)} = \gamma_v \partial_t \left[ (\mathbf{V}\times \mathbf{B}_0)\times  \mathbf{B}_0 \right]+ \gamma_P  \partial_t \mathbf{B}_0
   + \gamma_Q  \partial^2_{t^2} \mathbf{E}_0
\end{equation}
for a neutral particle moving with speed $\mathbf{V}$. The first force seems related by a Lorentz-transformation applied to  Eq.~(\ref{Fa}) from the co-moving frame to the laboratory frame, with again no further constraints on $\gamma_{v}$. We will show it to exist including the full time-derivative. Surprisingly, a similar term involving two electric fields does not appear but could be argued to exist on the basis of a  ``dual" symmetry in electric and magnetic fields, as discussed in Refs.~\cite{barnettloudon} and \cite{bliokh}. For the second force to be symmetry-allowed, $\gamma_P$ must be both charge-odd and parity-odd. Charge-oddness in neutral particles can be facilitated by the Faraday effect whose coupling constant, the Verdet constant, is charge-odd. Rotatory power induced by a chiral structure inside the particle provides an optical material constant that is parity-odd. Thus,  $\gamma_P$  could be a ''magneto-chiral" material constant associated with the mixture of both optical effects. This term is not predicted by the classical Maxwell equations, but was recently predicted to occur for a chiral molecule in a magnetic field \cite{Donaire} coupled to the quantum vacuum. Finally, the last term in Eq.~(\ref{Fb}) has just the restriction of charge-oddness of $\gamma_Q $. We will show that it emerges from the classical Maxwell equations for any finite charge distribution, which is a notion that goes beyond the usual description in terms on $\epsilon $ and $\mu$.

The Abraham-Minkowski controversy entered the closely related study of the electromagnetic angular momentum and its time-derivative, the torque. This was especially triggered by the experiment of Kristensen and Woerdman \cite{torqueexp} that interpreted their observed torques by a transfer of angular momentum equal to $1.03 \pm 0.05 \, \hbar$ per photon to a microwave antenna suspended in a circular waveguide, and which was seen to be independent on the index of refraction. This was argued to be consistent with the Minkowski version according to which the torque is equal to the full time derivative of the angular momentum  $ \int d^3\mathbf{r}\, \mathbf{R} \times (\mathbf{D}\times \mathbf{B})$. Similarly we can propose a number of possibilities for the (parity-even) electromagnetic torque $\mathbf{N}_A= \int d^3\mathbf{r}\, \mathbf{r} \times \mathbf{F}_A$ on a small particle at position $\mathbf{R}$,
 \begin{eqnarray}\label{Torqueb}
    \mathbf{N}_A
    &=& \Gamma_J \mathbf{R}\times \partial_t \left( (\mathbf{V} \times \mathbf{B}_0)\times\mathbf{ B}_0 \right)  + \Gamma_A \mathbf{R} \times \partial_t \left( \mathbf{E}_0\times  \mathbf{B}_0\right)\nonumber \\
    &&+    \Gamma_Q \partial_t \mathbf{B}_0+ \Gamma_P \partial_t \left( \mathbf{E}_0\times  \mathbf{B}_0\right) \nonumber \\
    &&+ \Gamma_v \mathbf{V}\times \left(\mathbf{E}_0 \times \mathbf{B}_0\right)
\end{eqnarray}
The first two terms depend explicitly on the position $\mathbf{R}$ of the particle, and  express the \emph{extrinsic }torque exerted by the Abraham forces in Eqs.~(\ref{Fa1}) and (\ref{Fb}). The last three terms are  related to  intrinsic angular momentum, and independent on the chosen origin $\mathbf{R}$ \cite{inex}. The Lenz law in classical electrodynamics, and contained in the internal canonical angular momentum in quantum mechanics, is actually a manifestation of the third term, with $\Gamma_Q$ charge-odd. The fourth term can only exist when $\Gamma_P$ is parity-odd, that is in chiral matter. It is well-known that a  plane wave with circular polarization $\sigma = \pm$  carries a intrinsic ``spin" angular momentum longitudinal to the propagation direction \cite{bliokh}, usually associated with the Poynting vector $\mathbf{S}=\mathbf{E} \times \mathbf{B}/4 \pi $. The transfer of this angular momentum  to dielectric matter has been discussed  \cite{opticsexpress2005,loudon} with subtle roles of surfaces. A torque is also known to exist if different circular polarizations are absorbed differently (circular dichroism) by a chiral particle \cite{expAMabs1,expAMabs2,bliokh}. At very low frequencies optical absorption is negligible. The optical equivalent of the torque proportional to $\Gamma_P$ corresponds to an angular momentum induced by the matter proportional to the Poynting vector $\mathbf{S}$, \emph{independent} on both polarization and absorption. For slowly varying external fields, this torque emerges as a genuine ``Abraham phenomenon", the torque-equivalent of Eq.~(\ref{Fa}). We will show that this torque exists and that $\Gamma_P$ is exactly equal to the material constant associated with rotatory power. The last term involves the time dependence of $\mathbf{R}$, with $\mathbf{V}=d\mathbf{R}/dt$ the velocity of the particle, without any time-dependence of the electromagnetic fields. We will show that this torque also exists, with $\Gamma_v$ exactly equal to the static electric polarizability.

This paper explores different kinds of Abraham-type forces on a neutral collection of electromagnetic charges (a ``molecule") and estimates orders of magnitude. The second part addresses the Abraham force on a molecule within the formalism of QED. This work contains basically known physics, yet put in a new context of Abraham forces. The comparison of quantum and classical approach also provides a fresh view on what is in the classical theory often considered as ``subtle" or ``controversial". Radiation forces near atomic resonances were first discussed in Ref.~\cite{loudonradiative}, both classically and quantum-mechanically.

\section{Forces on matter in macroscopic Maxwell equations}

We start with the classical, microscopic Maxwell equations (in Gaussian units),

\begin{eqnarray}\label{MMa}
   \nonumber \nabla\cdot \mathbf{E} &=& 4\pi \rho_i \\
   \nonumber  \nabla\cdot \mathbf{B} &=& 0 \\ \nonumber
  \nabla \times \mathbf{E } &=& - \frac{1}{c_0}\partial_t \mathbf{B}  \\
  \nabla \times \mathbf{B}  &=&  \frac{1}{c_0}\partial_t \mathbf{E} + \frac{4\pi}{c_0} \mathbf{J}_i
\end{eqnarray}
In the classical microscopic picture the charge density and the current density inside atoms and molecules can be represented by a sum of moving point charges. Let the index $n$ refer to these internal charges, and the index $N$  to the $N^{th}$ molecule, assumed neutral.
\begin{eqnarray}
\rho (\mathbf{r},t) &=& \sum_{N,n} q_{N,n} \delta \left(\mathbf{r}- \mathbf{R}_N -\mathbf{x}^{(N,n)} \right) \nonumber \\
\mathbf{J} (\mathbf{r},t) &=& \sum_{N,n} q_{N,n} \left(\mathbf{V}_N +  \mathbf{v}^{(N,n)}\right) \delta \left(\mathbf{r}- \mathbf{R}_N -\mathbf{x}^{(N, n)} \right) \nonumber \\
\end{eqnarray}
 where we defined the center of mass $\mathbf{R}_N(t)$ of object $N$ and the internal coordinate $\mathbf{x}^{(N,n)}(t)$ of the $n^{th}$ charge. Both can be time-dependent and $\mathbf{V}_N= d\mathbf{R}_N/dt$ and $\mathbf{v}^{(N,n)}=d\mathbf{x}^{(N,n)}/dt$. To go from a microscopic towards a  macroscopic picture  an expansion is carried out in $\mathbf{x}^{(N, n)}$ until a certain order. It is customary to write  $\rho_i = - \nabla\cdot \mathbf{P}$ for the microscopic  charge density and  $\mathbf{J}_i= \partial_t \mathbf{P} + c_0\nabla\times \mathbf{M}$ for the the induced current density. Upon keeping only terms quadratic in $\mathbf{x}^{(N,n)}$, assuming total charge $\sum_{n} q_{N,n}=0$ for each molecule $N$,  as well as velocity $\mathbf{V}_N = 0$, the local, macroscopic polarization density $\mathbf{P}$ and magnetization density $\mathbf{M}$ are then given by,
 \begin{eqnarray}\label{pm2}
   \mathbf{P}(\mathbf{r},t) &=&\sum_N \mathbf{d }_N(t)\delta(\mathbf{r}-\mathbf{R}_N) - \sum_N\mathbf{Q}_N(t)\cdot \nabla \delta(\mathbf{r}-\mathbf{R}_N)    \nonumber \\
   \mathbf{M}(\mathbf{r},t) &=&   \sum_N \mathbf{m}_N(t) \delta(\mathbf{r}-\mathbf{R}_N)
  \end{eqnarray}
In this equation the internal distribution of charges $(q_{N,n},\mathbf{x}^{(N,n)} )$  defines the electric dipole moment $\mathbf{d}_N= \sum_n q_{N,n}\mathbf{x}^{(n)}$, as well as  $\mathbf{Q}_N = \frac{1}{2} \sum_n q_{N,n} \mathbf{x}^{(N,n)}\mathbf{x}^{(N,n)}$. The trace of the tensor $\mathbf{Q}_N$  describes a finite spatial separation of positive and negative charges in object $N$, the traceless part is equal to its quadrupole moment (see Ref.~\cite{Jackson} for a discussion). Finally, the magnetic dipole moment is $\mathbf{m}_N= (1/2c_0) \sum_n q_{N,n}\mathbf{x}^{(N,n)} \times
  \mathbf{v}^{(N,n)} $. A finite velocity $\mathbf{V}_N$ of the molecule will be considered in the next section.

 The Lorentz force density can be straightforwardly written as
\begin{eqnarray}\label{Lorentz}
 &&\sum_{N,n} m_{N,n}\delta\left(\mathbf{r}- \mathbf{R}_N - \mathbf{x}^{(N,n)} \right)\frac{d^2\left(\mathbf{R}_N +\mathbf{x}^{(N,n)}\right)}{dt^2} \nonumber \\
 &=& \sum_N  M_N\frac{d\mathbf{V}_N}{dt} \delta(\mathbf{r}-\mathbf{R}_N) \nonumber \\
  && - \nabla \cdot \sum_N\delta(\mathbf{r}-\mathbf{R}_N) \sum_n m_{N,n}
 {\mathbf{x}}^{(N,n)} \frac{d\mathbf{v}^{(N,n)}}{dt} \nonumber \\
 &\equiv& \rho(\mathbf{r}) \frac{d\mathbf{V}(\mathbf{r})}{dt} -  \nabla\cdot\mathbf{ X}(\mathbf{r}) :=  \rho_i \mathbf{E} + \frac{1}{c_0}\mathbf{J}_i \times \mathbf{B }
\end{eqnarray}
The fields $\mathbf{E}$ and $\mathbf{B}$ are here still the total fields, both the externally applied fields and any possible field generated by internal charges and currents. After some vector calculus with no further assumptions, the force density on the left hand side  can be transformed into one that transfers momentum  from  the  electromagnetic field to the particle,
\begin{eqnarray}\label{mm1}
    \nonumber\rho\frac{d\mathbf{V}}{dt} &-&  \nabla\cdot\mathbf{ X} = \mathbf{f}_P  + \mathbf{f}_C \\
    &-&\nabla\cdot \left(\mathbf{PE} -\mathbf{BM} - \frac{\mathbf{1}}{2} (\mathbf{P}\cdot \mathbf{E} - \mathbf{B}\cdot \mathbf{M}\right)
\end{eqnarray}
In this formula, we identify a force density exerted by the electromagnetic field on the matter, and caused by induced currents and charges given by
\begin{equation}\label{fM}
   \mathbf{f}_C = -\frac{1}{2}\nabla(\mathbf{P}\cdot \mathbf{E}) - \frac{1}{2}\nabla(\mathbf{M}\cdot \mathbf{B}) + P_m \nabla E_m + M_m \nabla B_m
\end{equation}
involving only spatial derivatives. If the particles are surrounded by vacuum where $\mathbf{P}$ and $\mathbf{M}$ vanish rigourously, the force density $\mathbf{f}_c$ vanishes outside any particle, and the first two full derivatives disappear when performing the volume integral over any particle, and produce no net force. The last two terms provide the text-book forces on  electric and magnetic dipole.  The second force density contains only a time derivative,
\begin{equation}\label{fE}
   \mathbf{f}_P = \frac{1}{c_0}\partial_t \left( \mathbf{P}\times \mathbf{B}\right)
\end{equation}
Apart from the two force densities $\mathbf{f}_P$ and $\mathbf{f}_C$,  Eq.~(\ref{mm1}) features the  divergence of a stress tensor $\nabla\cdot\mathbf{ U}$. This tensor $\mathbf{U}$ is in general not symmetric, which physically implies that it contributes to the radiative torque on the matter, as discussed later in this work. If we assume vacuum outside the volume where the matter resides, $-\nabla\cdot \mathbf{U}$ does not contribute to the radiative force since its volume integral vanishes, and the force on the particle is  just the sum of $\mathbf{F}_C$ and $\mathbf{F}_P$. If we ignore $\mathbf{M}$, the total optical force  is
\begin{equation}\label{totalforce}
   \mathbf{F}_P+\mathbf{F}_C= \int d^3\mathbf{r} \left[ P_m\nabla E_m + \frac{1}{c_0}\partial_t (\mathbf{P}\times\mathbf{ B})\right]
\end{equation}
whose different derivations have been discussed by Barnett and Loudon \cite{forceondiel}.
It is remarkable that the force $\int d^3\mathbf{r}\,  \mathbf{f}_C $ is symmetric in electric and magnetic parts, unlike the 
force $\int d^3\mathbf{r}\,  \mathbf{f}_P$, which features only the polarization and not the magnetization. To have such symmetry one would need an additional force density
\begin{equation}\label{fE2}
   \mathbf{f}_M = \frac{1}{c_0}\partial_t \left( \mathbf{E}\times \mathbf{M}\right) \, ?
\end{equation}
For this term to exist we are obliged to extend the standard Maxwell equations to a dual electromagnetic theory \cite{dual}. This proposed force, including an argument  why it should not be proportional to the ''more symmetrical" form  $\partial_t \left( \mathbf{M}\times \mathbf{E}\right) $  will be discussed  later when we address the force on a moving molecule.  In vacuum ''dual" symmetry is defined as the symmetry upon changing $\mathbf{E} \rightarrow \mathbf{B}$, and $\mathbf{B} \rightarrow -\mathbf{E}$. In the presence of $\mathbf{P}$ and $\mathbf{M}$, such symmetry is less evident and Maxwell's equation have to be revised. 


The force  $ \mathbf{F}_P=\int d^3\mathbf{r} \, \mathbf{f}_P$ is often identified as \emph{the} Abraham-like force as the momentum $\int d^3\mathbf{r}\mathbf{P}\times \mathbf{B}/c_0$  seems to add up  to the momentum $M\mathbf{V}$ of the matter. In this expression, the time-derivatives are clearly symmetric on  $\mathbf{E}$ and $\mathbf{B}$, and would correspond to $\gamma_E= \gamma_B$ in Eq.~(\ref{Fa}), their common value being determined by the relation between polarization and electric field, typically the static polarizability. For electromagnetic waves at high frequencies $\mathbf{P}(t)$ and $\mathbf{B}(t)$ oscillate at the same frequency, and the force $\mathbf{F}_P$ tends to cycle-average out, and thus becomes hard to measure.

The Abraham force can easily be understood from the Lorentz force exerted on induced charges and currents when we neglect the magnetic moment density $\mathbf{M}$. Upon inserting $\mathbf{J}_i= \partial_t \mathbf{P}$, the magnetic force density $\mathbf{J}_i \times\mathbf{ B}/c_0$ gives $\mathbf{F}= \int d^3 \mathbf{r} c_0^{-1}(\partial_t \mathbf{P} )\times\mathbf{ B} $, where the integral extends over the particle. The electric force is $\mathbf{F}=  \int d^3 \mathbf{r} \rho_i \mathbf{E} $, and with the induced charge density  $\rho_i= -\nabla\cdot \mathbf{P} $ this can be rewritten, after one integration by parts and  using the $3^{th}$ Maxwell equation, as
\begin{equation}\nonumber
    \int d^3 \mathbf{r}  -E_i \nabla \cdot \mathbf{P}   =  \frac{1}{c_0}\int d^3 \mathbf{r}  (\mathbf{P}\times \partial_t \mathbf{B })_i  + \int d^3 \mathbf{r}  P_m\partial_i E_m
\end{equation}
The second term on the right hand side is just $\mathbf{F}_C$. Even in the presence of an external homogeneous electric field, it can contribute, as shown in the next section.

We note that, in addition to Eq.~(\ref{mm1}), a second equation for momentum conservation can be written down. For $\mathbf{D} = \mathbf{E} + 4 \pi \mathbf{P} $ and $\mathbf{H }= \mathbf{B} - 4\pi \mathbf{M}$, it follows from Maxwell's equations (\ref{MMa}),
\begin{eqnarray}\label{mm2}
   \nonumber \frac{1}{4\pi c_0}\partial_t \left( \mathbf{E}\times \mathbf{B}\right)  &=&  -\mathbf{f}_P -\mathbf{f}_C  \\
     \nonumber  +\frac{1}{4\pi }\nabla &\cdot& \left\{\mathbf{DE} +\mathbf{BH} - \frac{\mathbf{1}}{2} \left(\mathbf{D}\cdot \mathbf{E} + \mathbf{B}\cdot \mathbf{H}\right)\right\}\\
\end{eqnarray}
In this equation, the separation between the different terms is not uniquely defined and the aforementioned different versions of electromagnetic momentum move either gradient terms contained in $\mathbf{f}_C$ to the stress tensor $\nabla\cdot (\cdots)$ or move time-derivatives from the Abraham force $\mathbf{f}_A$ to the momentum density on the left hand side \cite{Brevik}. Written in our format above, we have separated the force density $\mathbf{f}_C +\mathbf{f}_P$  as in Eq.~(\ref{mm1}), with opposite sign so that Newton's third law is manifestly obeyed. In this version, the electromagnetic momentum density is identified as $\mathbf{K}= (4\pi c_0)^{-1}\mathbf{E} \times \mathbf{B} $, which is the version derived by Nelson \cite{Nelson} in a more general approach. Upon adding (\ref{mm1}) and (\ref{mm2})  we establish ``total" momentum conservation

\begin{eqnarray}\label{mm3}
   \nonumber \partial_t \mathbf{K} & + & \rho\frac{d\mathbf{V}}{dt} - \nabla\cdot \mathbf{X} =  \\
 \nonumber    &+& \frac{1}{4\pi }\nabla \cdot  \left\{\mathbf{EE} +\mathbf{BB} - \frac{\mathbf{1}}{2} \left(\mathbf{E}\cdot \mathbf{E} + \mathbf{B}\cdot \mathbf{B}\right)\right\} \\
\end{eqnarray}
Quite conveniently, the momentum stress tensor in this equation is fully symmetric, but does no longer necessarily vanish outside the matter, as was the case  in Eq.~(\ref{mm1}). The symmetry indicates conservation of the sum of  the angular momenta of  electromagnetic field, the center of mass of the particles and the internal angular momentum.  Equation~(\ref{mm3}) is particularly useful in stationary situations and/or at high frequencies, i.e. for scattering of a plane wave by a small particle or atom.  In this case the time-derivative of field momentum-density vanishes, and the total force can be found as either the integral of the stress tensor over the surface of the particle or, even simpler, over a surface in the far field, where the fields are described by the on-shell cross-section.  In the presence of slowly varying homogeneous external fields, typically true in the context of the Abraham force, the force $ (4\pi c_0)^{-1} \int_V d^3 \mathbf{r} \partial_t(\mathbf{E}\times \mathbf{B})$ on the lefthand side of Eq.~(\ref{mm3}) exerted on any volume $V$ outside the particle must be counterbalanced on the righthand side  by the force created on the surface of $V$ by  \emph{time-induced} electric and magnetic fields in the stress tensor, and without any impact on the particle. As a result, Eq.~(\ref{mm1}) is far more appropriate to use, and the electromagnetic force on the particle rigorously given by the volume integral of $\mathbf{f}_P+\mathbf{f}_C$ that only extends over the object.

\section{Forces by slowly varying classical fields}

\subsubsection{Electromagnetic force on molecule at rest}

  In the following we consider one small, neutral particle consisting of a number point charges, and refer to this as a ``molecule",  and ignore the label particle  $N$ introduced in the previous section. Maxwell's equations \emph{in free space} imply that for slowly-varying electromagnetic fields $\mathbf{E}_0(t)$ and $\mathbf{B}_0(t)$,
  \begin{eqnarray} \label{linEB}
   \mathbf{E}(\mathbf{r},t) &=& \mathbf{E}_0(t) + \frac{1}{2c_0} \mathbf{r} \times \partial_t \mathbf{B}_0 + \frac{1}{10c_0^2}\left(
   2 \mathbf{r}^2 - \mathbf{r} \mathbf{r}\right) \cdot \partial^2_t \mathbf{E}_0 \nonumber \\
   \mathbf{B}(\mathbf{r},t) &=& \mathbf{B}_0(t) - \frac{1}{2c_0} \mathbf{r }\times \partial_t \mathbf{E}_0
   + \frac{1}{10c_0^2}\left(
   2 \mathbf{r}^2 - \mathbf{r} \mathbf{r}\right) \cdot \partial^2_t \mathbf{B}_0
   \nonumber \\
 \end{eqnarray}
 We define the location $\mathbf{r}=0$ as the center of the ``homogeneous" fields $\mathbf{E}_0$ and $\mathbf{B}_0$ that in any real experiment are never homogeneous.
  We emphasize that the equations above neglect contributions from the matter, via its polarization density $\mathbf{P}$ and magnetization density $\mathbf{M}$, and we shall assume that induced fields are included in the constitutive equations relating $\mathbf{P}, \, \mathbf{M}$ to $\mathbf{E}, \, \mathbf{B}$. The time-dependence of the electric field induces a magnetic field $\mathbf{B}_i(\mathbf{r},t) = \frac{1}{2c_0 } \partial_t \mathbf{E}_0 \times \mathbf{r}$ and vice versa. Higher order  terms can be neglected as long as the typical time-dependence $\partial_t \sim \omega$ of the fields is sufficiently slow so that $\omega r/c_0 \ll 1$.  On distances of $r = 1 $ cm, this implies $\omega < 30 $ GHz.  In our experiments is $\omega/ 2\pi  \approx $ kH. As long as these frequencies are small compared to the resonant frequencies  $\omega_0 \sim 10^{14} $ Hz, the matter will follow  the external fields quasi-statically. The linear dependence on $\mathbf{r}$  of the two slowly varying electromagnetic fields in Eqs.~(\ref{linEB}) actually contributes to the standard Abraham force $\mathbf{F}_P$ as we see below.
  The total electromagnetic force is given by

  \begin{equation}\nonumber
    \sum_n m_n \frac{d^2\mathbf{r}^{(n)}}{dt^2} = M\frac{d\mathbf{V}}{dt} = \int d^3\mathbf{r} (\mathbf{f}_P + \mathbf{f}_{CE} + \mathbf{f}_{CB})
  \end{equation}
  with $\mathbf{f}_P$ given by Eq.~(\ref{fE}) and the two others by  Eq.~(\ref{fM}). The integral extends over the molecule's volume. We insert for the induced electric dipole moment $\mathbf{d}(\mathbf{r},t) = \alpha(0) \mathbf{E}(\mathbf{r},t)$, with $\alpha(0)$ the static electric polarizability.  We assume pure vacuum outside the particle so that the terms $\nabla\cdot \mathbf{U}$ and  $\nabla\cdot \mathbf{X}$ in Eq.~(\ref{mm1}) vanish in the integral. We adopt the possibility of a permanent magnetic dipole moment $\mathbf{m}$, but exclude for simplicity any induced magnetic dipole. Any force determined by $\mathbf{Q}$ is only non-zero if the second-order terms in  Eq.~(\ref{linEB}) are retained, and will be dealt with below.  The first contribution to $\mathbf{F}_P$ defined in Eq.~(\ref{fE}) is given by

 \begin{eqnarray}\label{fball}
    \mathbf{F}_P &=& \frac{\alpha(0))}{c_0} \partial_t \left( \mathbf{E}_0 \times \mathbf{B}_0\right) \nonumber \\
    &+& \frac{\alpha(0)}{2c_0^2} \partial_t \left[ (\mathbf{R}\times \partial_t\mathbf{B}_0 ) \times \mathbf{B}_0 \right]\nonumber \\
   &+&   \frac{\alpha(0)}{2c_0^2} \partial_t \left[ (\mathbf{R}\times \partial_t\mathbf{E}_0 ) \times \mathbf{E}_0 \right]
 \end{eqnarray}
 The second contribution stems from  $\mathbf{f}_{CE,i} = P_m\partial_i E_m$ and integrates to,
 \begin{eqnarray}\label{FEall}
    \mathbf{F}_{CE} =  -\frac{\alpha(0)}{2c_0}  \left( \mathbf{E}_0\times \partial_t\mathbf{B}_0\right) + \frac{\alpha(0)}{4c_0^2} \partial_t \mathbf{B}_0 \times( \mathbf{R} \times \partial_t\mathbf{B}_0) \nonumber \\
    - \frac{\alpha(0)}{10c_0^2} \left[ \mathbf{E}_0(\mathbf{R}\cdot \partial^2_t\mathbf{E}_0) +  \partial^2_t\mathbf{E}_0(\mathbf{R}\cdot \mathbf{E}_0) - 4\mathbf{R}
    (\mathbf{E}_0\cdot \partial^2_t\mathbf{E}_0) \right] \nonumber \\
 \end{eqnarray}
The force $\mathbf{f}_{CM,i} = M_m\partial_i B_m$ gives us,
  \begin{eqnarray}
    \mathbf{F}_{CM} = \frac{1}{2c_0} \mathbf{ m }\times \partial_t\mathbf{E}_0 \nonumber
 \end{eqnarray}
Finally, the Abraham force induced by $\mathbf{Q}$ can be best obtained by realizing that the electric force density  is just $f_i= - {E }_i\partial_j P_j
= E_i Q_{kj}\partial_k\partial_j \delta(\mathbf{r}-\mathbf{R})$ while the magnetic force vanishes since  $\partial_t \mathbf{P}=0 $ for the quadrupole contribution to $\mathbf{P}$. Inserting the quadratic terms of $\mathbf{E}(\mathbf{r},t)$ leads to
\begin{equation}\label{FQ}
    \mathbf{F}_Q = \frac{1}{5c^2_0} \left(2 \mathrm{Tr} \, \mathbf{Q} \partial^2_t \mathbf{E}_0 -\mathbf{ Q}\cdot \partial^2_t \mathbf{E}_0  \right)
\end{equation}
The total force $\mathbf{F}_P+\mathbf{F}_{CE}+ \mathbf{F}_{CM}+   \mathbf{F}_Q $ can be rewritten as a sum of 1) an Abraham-type force $\mathbf{F}_A$ involving time-derivatives of both $\mathbf{E}_0$ and $\mathbf{B}_0$, 2) a force $\mathbf{F}_{MQ}$ stemming from the permanent magnetic dipole $\mathbf{m}$ and the electric quadrupole $\mathbf{Q}$,   3) a force $\mathbf{F}_R$ that depends linearly on $\mathbf{R}$ in all directions and can be both attractive or repulsive, and 4) a velocity-dependent force $\mathbf{F}_V$ linear in $d\mathbf{R}/dt= \mathbf{V}/c_0$, that will be discussed in the next section. The first terms of $\mathbf{F}_P+\mathbf{F}_{CE}$ add up to an Abraham-type force,

\begin{equation}\label{Fa1}
    \mathbf{F}_A = \frac{\alpha(0)-\frac{1}{2}\chi(0)}{c_0} \left(\partial_t \mathbf{E}_0 \times \mathbf{B}_0 \right)
    + \frac{1}{2}\frac{\alpha(0)}{c_0} \left( \mathbf{E}_0 \times \partial_t \mathbf{B}_0 \right)
\end{equation}
For completeness we have re-introduced the magnetic susceptibility in this expression, that comes in via the force $\int d^3 \mathbf{r}  \, M_m \nabla B_m$ in Eq.~(\ref{fM}) using the induced magnetic moment density $\mathbf{M} = \chi(0) \mathbf{B}$.
The force $\mathbf{F}_Q$   grouped with $\mathbf{F}_{CM}$ is,
\begin{eqnarray}\label{FMQ}
    \mathbf{F}_{MQ} =    \frac{1}{5c_0^2}\left(2\mathrm{Tr }\mathbf{Q} - \mathbf{Q}\right) \cdot  \partial^2_t \mathbf{E}_0 + \frac{1}{2c_0} \mathbf{ m }\times \partial_t\mathbf{E}_0
\end{eqnarray}
For $\mathbf{F}_R$ we find
 \begin{eqnarray}\label{FRmagnetic}
   \mathbf{F}_R^{(M)} &=& \frac{\alpha(0)}{2c_0^2}   (\mathbf{R }\times \partial_t^2 \mathbf{B}_0) \times \mathbf{B}_0   \nonumber \\
&& \ \ + \frac{\alpha(0)}{4c_0^2} (\mathbf{R}\times \partial_t\mathbf{B}_0) \times \partial_t\mathbf{B}_0
 \end{eqnarray}
 associated with the magnetic field, and
 \begin{eqnarray}\label{FRelectric}
  && \mathbf{F}_R^{(E)} =  \frac{\alpha(0)}{2c_0^2} \partial_t \left[ (\mathbf{R}\times \partial_t\mathbf{E}_0 ) \times \mathbf{E}_0 \right] \nonumber \\
 &&- \frac{\alpha(0)}{10c_0^2} \left[ \mathbf{E}_0(\mathbf{R}\cdot \partial^2_t\mathbf{E}_0) +  \partial^2_t\mathbf{E}_0(\mathbf{R}\cdot \mathbf{E}_0) - 4\mathbf{R}
    (\mathbf{E}_0\cdot \partial^2_t\mathbf{E}_0) \right] \nonumber \\
 \end{eqnarray}
 for the electric field.
Equation~(\ref{Fa1}) denotes an an Abraham force that is \emph{asymmetric} in the time-derivatives by exactly a factor of $2$. This factor is modified  by any finite magnetic permeability $\chi(0)$. We found a similar factor of two difference between gradient force and Abraham force elsewhere in a different discussion of radiation trapping of atoms \cite{AbMiBarnett}. The force~(\ref{FMQ}) relies, like the Abraham term, entirely on internal coordinates, and depends only on the time-dependent electric field. It is non-zero for either a permanent magnetic moment (last term) or for a permanent charge separation (first term). It is small but interesting since it couples the electric field directly to the moments $\mathbf{Q}$ and $\mathbf{m}$. For instance, we can consider the CO$_2$ molecule which has a permanent electric quadrupole moment $Q= -1.4 \cdot 10^{-39}$ Cm$^2$. A homogeneous, static electric field  creates nor a force linear in the field since the total charge is zero, neither one quadratic in the field since the gradient vanishes. The time-dependence creates an electric field that varies quadratically over the particle. The tensor  $\mathbf{Q}$ is associated with a time-independent, charge distribution inside, which results in a net force. The effect is obviously small. We estimate for $E_0= 10^5 $ V/m oscillating at $\omega = 3\cdot 10^4$ Hz, $F_Q \approx 1.5\cdot 10^{-42} $ N. For comparison, the standard Abraham force under the same circumstances and at $B_0=1$ T equals $F_A= 6 \cdot 10^{-32}$ N and $F_R \approx  \alpha(0)\omega^2B_0^2 R /2c_0^2 \approx 10^{-36}$ N at a distance of $R=1 $ mm. The expansion in Eqs.~(\ref{linEB}) applies as long as $\omega a/c_0 \ll 1$ with $a$  the particle size, so especially the expression for $\mathbf{F}_{MQ}$ should be valid at much higher frequencies. Nevertheless it is an experimental challenge to detect such rapid movements time-resolved.

\subsubsection{Abraham Force on a slab}

In this section we show that the  Abraham force found in Eq.~(\ref{Fa1}) for a neutral molecule, with the property $ \gamma_E = 2\gamma_B$, does not apply to the slab geometry. The radiation pressure at optical frequencies on a dielectric slab was discussed by Kim etal. \cite{Kim} using a similar formalism, but did not address the symmetry of the time-derivatives. In a macroscopic picture we write $4\pi \mathbf{P} = (\epsilon -1) \mathbf{E}(\mathbf{r}) $ in Eq.~(\ref{pm2}).
We evaluate the total optical force on a dielectric slab with thickness $\Delta z= d$ and assume transactionally symmetric in $xy$ (that is for a large surface $A$, with neglect of boundary effects).
For a homogeneous magnetic field $B_0(t)\mathbf{\hat{y}}$ in the $\hat{y}$-direction, and a homogeneous electric field $E_0(t)\mathbf{\hat{z}}$ across the slab,  the Abraham force is just

\begin{equation}\label{Fpslab}
\mathbf{F}_P = \frac{Ad}{4\pi c_0} (\varepsilon-1)\frac{d}{dt} (\mathbf{E}_0\times \mathbf{B}_0)
  \end{equation}
and is directed along the $\hat{x}$-direction parallel to the slab. In the force $\mathbf{F}_C$ in Eq.~(\ref{fM}) only the term $-E_n\nabla P_n$ survives the integral over the slab. In view of the translational $xy$-symmetry, the electric field induced by the time-dependent magnetic field can only depend on $z$ and the 3th Maxwell equation gives  $\partial_zE_x= -(1/c_0) dB_0/dt$. It easily follows that $\mathbf{F}_C=0$, so that the Abraham force  given by Eq.~(\ref{Fpslab})  is just the total force on the slab. Hence we find here $\gamma_E=\gamma_B$ in  Eq.~(\ref{Fa1}), unlike the outcome~(\ref{Fa1}) for a molecule. This result is in agreement with the experiment in Ref.~\cite{Geert2}.

\subsubsection{Abraham Force on a sphere}

In this section we solve Maxwell's equations for a homogeneous sphere of radius $a$ with static permeabilities  $\varepsilon$ and $\mu$, exposed to slowly varying homogeneous,orthogonal fields $\mathbf{E}_0$ and $\mathbf{B}_0$. We shall find the Abraham force linear in both, and compare to Eq.~(\ref{Fa1}) derived for a ``classical, neutral  molecule". An Abraham-Minkowski controversy exists on how this force depends on $\varepsilon$ and $\mu$ \cite{Brevik,Nelson}. It is well known \cite{Jackson} that the fields inside the sphere are homogeneous and equal to $\mathbf{E}_{in}= 3\mathbf{E}_0/(\varepsilon+2)$ and $\mathbf{B}_{in}= 3\mu \mathbf{B}_0/(\mu+2)$ . If we include the induced fields we find, inside the sphere,
\begin{eqnarray}
 \nonumber  \mathbf{E} &=&  \mathbf{E}_{in} -\frac{1}{2c_0 } \partial_t \mathbf{B}_{in} \times \mathbf{r}  \\
 \nonumber  \mathbf{B}  &=& \mathbf{B}_{in}+ \frac{\varepsilon\mu}{2c_0 } \partial_t \mathbf{E}_{in} \times \mathbf{r}\\
\end{eqnarray}
This time, and contrary to Eq.~(\ref{linEB}), this is the simple but rigorous solution, with all locally induced fields included. The small parameter  is still  $a / c_0\Delta t $, where $\Delta t$ is the time scale on which the field change. Since the polarization density is $\mathbf{P} = (\varepsilon-1)\mathbf{E}_{in}/4\pi$ we get
  \begin{equation}\nonumber
    \mathbf{F}_P= \frac{a^3}{3 c_0} (\varepsilon-1) \partial_t (\mathbf{E}_{in} \times \mathbf{B}_{in})
  \end{equation}
The force $\mathbf{F}_C$ can be evaluated straightforwardly ($\mathbf{M}= (1-1/\mu) \mathbf{B}_{in}/4\pi$)
   \begin{eqnarray*}
    \mathbf{F}_C= - \frac{a^3}{6 c_0} (\varepsilon-1)  \mathbf{E}_{in} \times (\partial_t \mathbf{B}_{in})
   \\  - \frac{a^3}{6 c_0} \varepsilon (\mu-1)   (\partial_t\mathbf{E}_{in}) \times \mathbf{B}_{in}
  \end{eqnarray*}
which brings us to the total Abraham force
\begin{eqnarray}
  \mathbf{F}_A &=& \frac{a^3}{3 c_0} \left(\frac{3}{2}\varepsilon-1 - \frac{1}{2}\varepsilon \mu \right)  (\partial_t\mathbf{E}_{in}) \times \mathbf{B}_{in}   \\
  &+&  \frac{a^3}{6 c_0} (\varepsilon-1)  \mathbf{E}_{in} \times \partial_t \mathbf{B}_{in}
\end{eqnarray}
For $\mu=1$  we find $\gamma_E=2\gamma_B$.  If we write the static polarizability of the sphere as a whole as $\alpha(0) = (\varepsilon-1)a^3 / (\varepsilon +2)$, we find $\gamma_E= \alpha(0)/c_0$ as was found for a molecule in  Eq.~(\ref{Fa1}). The magnetic force for $\mu\neq 1$ corresponds to the force $\mathbf{F}_M$ in Eq.~(\ref{fM}) for which $\gamma_B=0$ and
$\gamma_E=-(\mu -1)a^3 / 2c_0(\mu +2) = -\chi(0)/2c_0$, again the same as in Eq.~(\ref{Fa1}). We conclude that the Abraham force on a neutral molecule and a polarizable sphere are equivalent.

\subsubsection{Force on moving molecule}

In this section we consider a moving, small neutral particle containing charges, still referred to as a ``molecule",  and find the velocity-dependent force $\mathbf{F}_V$ exerted on the molecule by the electromagnetic field.
We emphasize that Eq.~(\ref{mm1}) is valid in all reference frames, co-moving or not. The explicit velocity dependence comes in only via the constitutive equations for $\mathbf{P}$ and $\mathbf{M}$ that take the form~(\ref{pm2}) in the co-moving frame. Hence in the laboratory frame where the particle moves with speed $\mathbf{V}$ we have
\begin{equation}\label{Pv}
    \mathbf{P}(\mathbf{r},t) = \alpha(0) \delta(\mathbf{r}-\mathbf{R}(t)) \left[ \mathbf{E} + \frac{\mathbf{V}}{c_0} \times \mathbf{B}\right] + \frac{\mathbf{V}}{c_0} \times \mathbf{M}
\end{equation}
since $\mathbf{E} + \frac{\mathbf{V}}{c_0} \times \mathbf{B}$ is just the electric field in the co-moving frame where $\mathbf{P}'= \alpha(0)\delta(\mathbf{r}-\mathbf{R}(t))\mathbf{E}' $, the second term stems from the Lorentz transformation of $\mathbf{D}$ and $\mathbf{H}$. We will ignore the effect of internal magnetization, and write
\begin{equation}\label{Mv}
\mathbf{M} = -\frac{1}{c_0}\mathbf{V}\times \mathbf{P} = -\frac{\alpha(0)}{c_0} \delta(\mathbf{r}-\mathbf{R}(t)) \mathbf{V} \times \mathbf{E}_0
\end{equation}
This contribution to the magnetization is equivalent to the ``R\"{o}ntgen current $\mathbf{J}= \nabla \times (\mathbf{P} \times \mathbf{V}) = \mathbf{V}\rho + (\mathbf{V}\cdot \nabla) \mathbf{P}$ discussed by Loudon etal, and essentially a Galilean transformation of the current density \cite{loudonmomentum}. The polarization density $\mathbf{P}=\mathbf{V} \times\mathbf{ M}/c_0$ corresponds to an induced charge density $\rho= \mathbf{V}\cdot \mathbf{J}/c_0^2$  used in Ref.~\cite{Aharonov} to find the Aharonov-Casher topological phase for a neutral, magnetic particle. If we substitute Eq.~(\ref{linEB})  for $\mathbf{B}$, we immediately get a magnetic dipole  force

\begin{equation}\label{FV1}
    \mathbf{F}_V^{(1)} = \int d^3\mathbf{r} M_m \nabla B_m = \frac{\alpha(0)}{2c_0^2} \partial_t \mathbf{E}_0 \times (\mathbf{V}\times \mathbf{E}_0)
\end{equation}
The time derivative in the force $\mathbf{F}_P$ already derived in Eq.~(\ref{fball}) generates two terms linear in velocity,
\begin{eqnarray*}
 \mathbf{F}_V^{(2)} &=& \frac{\alpha(0)}{2c_0^2} \left[ (\mathbf{V}\times \partial_t\mathbf{E}_0 ) \times \mathbf{E}_0 \right]    \\
  \, &+& \frac{\alpha(0)}{2c_0^2} \left[ (\mathbf{V}\times \partial_t\mathbf{B}_0 ) \times \mathbf{B}_0 \right]
\end{eqnarray*}
In the  force $\mathbf{F}_P$ in Eq.~(\ref{fball}) we must also acknowledge a contribution from the  polarization $\mathbf{d} = \alpha(0)\mathbf{V}/c_0 \times \mathbf{B}_0$ so that,
\begin{eqnarray*}
 \mathbf{F}_V^{(3)} &=& \frac{\alpha(0)}{c_0^2} \partial_t \left[ (\mathbf{V}\times \mathbf{B}_0 ) \times \mathbf{B}_0 \right]
\end{eqnarray*}
The electric dipole force achieves a velocity-dependent contribution,
\begin{eqnarray*}
 \mathbf{F}_V^{(4)} &=&   \int d^3\mathbf{r} P_m \nabla E_m  \nonumber \\
 &=& \int d^3\mathbf{r}  \alpha(0) \left(\frac{\mathbf{V}}{c_0}\times \mathbf{B}_0\right)_m \delta(\mathbf{r}-\mathbf{R})\nabla \left( \frac{1}{2c_0} \mathbf{r}\times \partial_t \mathbf{B}_0\right)_m \\
&=&  \frac{\alpha(0)}{2c_0^2} \left[ \partial_t\mathbf{B}_0\times (\mathbf{V}\times \mathbf{B}_0 )   \right]
\end{eqnarray*}
Upon adding up the four contributions, thereby using $\mathbf{a }\times (\mathbf{b}\times \mathbf{c}) +\mathbf{c }\times (\mathbf{a}\times \mathbf{b}) + \mathbf{b }\times (\mathbf{c}\times \mathbf{a})=0$, we arrive at
\begin{eqnarray} \label{FVall}
  \mathbf{F}_V &=& \frac{\alpha(0)}{c_0^2} \partial_t \left[ (\mathbf{V}\times \mathbf{B}_0) \times \mathbf{B}_0 \right] \nonumber \\
   &+&  \frac{\alpha(0)}{2c_0^2}  \mathbf{V}\times \left(\partial_t \mathbf{B}_0 \times \mathbf{B}_0 \right)\nonumber \\
   &+&  \frac{\alpha(0)}{2c_0^2}  \mathbf{V}\times \left(\partial_t \mathbf{E}_0 \times \mathbf{E}_0 \right)
\end{eqnarray}
The last two terms induce a slow Lorentz-type  deflection of the molecule if the fields themselves rotate. They are symmetric in electric and magnetic fields, but vanish if the direction of the fields is constant in time. The first term can be considered as a contribution of the magnetic field to the inertial ``mass tensor" of the particle, of order $\Delta M =  \alpha(0)B_0^2/c_0^2$, time-dependent if $\mathbf{B}_0$ varies with time, and different along and perpendicular to the field lines. If we include the polarization density $\mathbf{P} = \mathbf{V} \times \mathbf{M} /c_0= \chi(0)\mathbf{V}\times \mathbf{B}_0/c_0 $ imposed by Lorentz-invariance in Eq.~(\ref{Pv}), we find an additional term containing again the full time-derivative of two magnetic fields, but now proportional to $\chi(0)$. Hence,  $\Delta M =  (\alpha(0)+\chi(0)) B_0^2/c_0^2$. Somewhat surprisingly, since the Lorentz transformation of electromagnetic fields has been used explicitly,  no equivalent contribution exists from the electric field, which we could have expected noting that $\frac{1}{2} \alpha(0) \mathbf{E}_0^2 +\frac{1}{2} \chi(0)\mathbf{ B}_0^2$ is the extra electromagnetic energy stocked inside the molecule, and should be equivalent to a change in mass $\Delta M c_0^2$.  Technically, this absence stems from the fact that the Abraham force density $\partial_t (\mathbf{P}\times \mathbf{B})$ produces, in the laboratory frame, only the velocity-dependent force $\mathbf{F}^{(3)}_V$ found above. To generate the equivalent expression with the electric field $\mathbf{E}_0$ one would need an additional Abraham force density $\partial_t (\mathbf{E}\times \mathbf{M})/c_0 $ since the moving molecule is subject to a magnetization density $\mathbf{M}= -\mathbf{V} \times \mathbf{P}/c_0 = -\alpha(0) \mathbf{V} \times \mathbf{E}_0/c_0 $. This force was already proposed in Eq.~(\ref{totalforce}) on the basis of symmetry arguments in magnetic and electric components. The relation between the equivalence of mass and electromagnetic energy should be subject to a more profound study, using recent advances in dual extensions of Maxwell's equations \cite{dual}. Unfortunately the effect is extremely small. For Helium in a magnetic field of 1 Tesla  we estimate $\Delta M/M  = 3.4 \cdot 10 ^{-13}$. We can argue that a similar result holds for a dielectric sphere, and we estimate in SI units, $\Delta M /M =  3 \epsilon_0 B_0^2 (\epsilon(0)/\epsilon_0-1)/(\epsilon(0)/\epsilon_0+2)/\rho $, with $\rho$ the mass density. For water droplets in 10 T this yields $\Delta M /M\approx 10^{-12}$.

\subsubsection{Torque on a molecule}

The torque exerted on a molecule is defined as

\begin{equation}\label{torque}
    \mathbf{N}(t) = \int d^3 \mathbf{r} \, \, \mathbf{r}\times \mathbf{f} (\mathbf{r}, t)
\end{equation}
where the integral extends over its volume and depends in general on the origin defined by $\mathbf{r}=0$, and $\mathbf{f}$ is given by the right hand side of  Eq.~(\ref{mm1}). From this equation we get immediately that
\begin{eqnarray}\label{torque2}
 &&\frac{d}{dt} \left( M \mathbf{R} \times\mathbf{ V }+  \sum_n m_n \mathbf{r}^{(n)} \times  \mathbf{v}^{(n)} \right) = \mathbf{N}(t) \nonumber \\
 =&&  \int d^3 \mathbf{r} \,  \mathbf{r}\times (\mathbf{f}_P + \mathbf{f}_C ) +  \int d^3 \mathbf{r} \, (\mathbf{P}\times \mathbf{E} + \mathbf{M} \times \mathbf{B} )
\end{eqnarray}
The torque equals the rate of change of internal plus external angular momentum.
The first integral in the torque is essentially the torque corresponding the Abraham force discussed in a previous section. Most terms can be copied immediately as,

\begin{eqnarray}
 \mathbf{N}_1=  \mathbf{R} \times (\mathbf{F}_A + \mathbf{F}_R + \mathbf{F}_{CM} + \mathbf{F}_V )
\end{eqnarray}
 and is entirely external, i.e. zero if we choose $\mathbf{R}=0$. One integration by parts that has led to  $\mathbf{F}_P$ in Eq.~(\ref{fball}) has to be reconsidered for the torque, and will be discussed below. The quadrupole  $\mathbf{Q}$ also needs a special treatment due to the higher moment of the couple,
\begin{eqnarray}\label{lenztorque}
 \mathbf{N}_2=  \frac{d}{dt}\int d^3 \mathbf{r}\,  \mathbf{r} \times (-\mathbf{Q}\cdot \nabla \delta (\mathbf{r}-\mathbf{R}_N) ) \times \mathbf{B}_0 \nonumber \\
 = \sum_n \frac{q_n}{2c_0}
    \partial_t (\mathbf{x}^n\times (\mathbf{x}^n \times \mathbf{B}_0) )
\end{eqnarray}
This torque can be associated with the electric force that according to the Lenz induction law is proportional to the time-derivative of the enclosed magnetic flux. It is also equal to the contribution of the electromagnetic field $q_n \mathbf{x}_n \times \mathbf{A}(\mathbf{x}_n) /2c_0$ to the internal canonical angular momentum of the matter \cite{Cohen}, when expressed in the Coulomb gauge.

The second term in Eq.~(\ref{torque2}) is the textbook result for electric and magnetic moments \cite{Jackson}. If we exclude permanent moments, we see that it vanishes in isotropic matter where $\mathbf{P} \sim \mathbf{E}$ and $\mathbf{M}\sim \mathbf{B}$. We shall discuss a few special cases where it does not vanish. In general the presence of the fields $\mathbf{E}_0(t)$ and $\mathbf{B}_0(t)$ couple internal and external angular momentum which are no longer individually conserved. Let us first look at the terms that do not depend on $\mathbf{R}$ and that can be considered as  ``internal" torques, intrinsic to the molcule. For rotatory power it is well-know that $\mathbf{m}=\int d^3 \mathbf{r}\, \mathbf{M}= g(0) \partial_t \mathbf{E}_0/c_0$ and  $\mathbf{d}=\int d^3 \mathbf{r}\,\mathbf{P}= -g(0) \partial_t \mathbf{B}_0/c_0 $, with the same static rotatory power permeability  $g(0)$ appearing in both constitutive equations, and possibly non-zero when the molecule is chiral \cite{Craig}. This gives
\begin{eqnarray}\label{chiral}
    \mathbf{N}^{\mathrm{RP}}_3&= & \mathbf{d}\times \mathbf{E} + \mathbf{m} \times \mathbf{B}   \nonumber \\
    &=&  \frac{g(0)}{c_0} \partial_t\left( \mathbf{E}_0 \times \mathbf{B}_0 \right)
\end{eqnarray}
This is an unfamiliar contribution to the torque induced by chirality, and even a genuine contribution $ (g(0)/c_0)  \mathbf{E}_0 \times \mathbf{B}_0 $ to the internal angular momentum of the molecule. For a magnetic field oscillating at $\omega = 10^5$ Hz and $qx^2/2 \sim Q \approx  10^{-39}$ Cm$^2$ the Lenz torque is $N_2 \approx 10^{-34}$ J/T. For the chiral compound 2-octanol ($C_{8}H_{18}O$) we estimate
$g(0) = 3 \cdot  10^{-53}$ Cm$^3$/V \cite{Condon}, which implies for an electric field $E_0 = 10^5 $ V/m a chiral torque of $3 \cdot 10^{-43} $ J/T, which is indeed very small compared to the torque $\mathbf{N}_2$. Maybe in artificial meta-materials this torque induced by broken mirror-symmetry can be made observable .

For  \emph{stationary}, crossed electric and magnetic fields acting on a moving molecule, we insert $\mathbf{m} = \chi(0) (\mathbf{B}_0 -\mathbf{V} \times\mathbf{ E}_0/c_0) - \mathbf{V }\times \mathbf{d}/c_0$ and $\mathbf{d}= \alpha(0) (\mathbf{E}_0 + \mathbf{V} \times\mathbf{ B}_0/c_0) + \chi(0) \mathbf{V} \times \mathbf{B}_0/c_0$, it follows that

\begin{equation}\label{vEB}
    \mathbf{N}^{\mathrm{V}}_3    = -\frac{\alpha(0) + \chi(0) }{c_0} \mathbf{V} \times (\mathbf{E}_0 \times \mathbf{B}_0)
\end{equation}
or equivalently, the conservation of $\mathbf{J}+ (\alpha(0) + \chi(0) ) {\mathbf{R}} \times (\mathbf{E}_0 \times \mathbf{B}_0)/c_0$. This suggest that
$(\alpha(0) + \chi(0) )  (\mathbf{E}_0 \times \mathbf{B}_0)/c_0$ emerges as an "Abraham" momentum even when the electromagnetic fields are time-independent, and contributes to total angular momentum.
For a molecule rotating initially around the vector $\mathbf{E}_0 \times \mathbf{B}_0$, this effect induces the rotation axis to start rotating   slowly itself.
For a typical static polarizability $\alpha(0) =   10 ^{-39}$ C$^2$m$^2$/V  we find a torque $N \approx  10 ^{-34} $ J/T  for a velocity of $10$ km/s, of comparable to the Lenz torque $N_2$.

A special case that covers the three torques $\mathbf{\mathbf{N}}_1$, $\mathbf{N}_2$ and $\mathbf{N}_3$  is the one of a moving molecule in the absence of an electric field, but subject to a time-dependent magnetic field with constant $z$-orientation. In this case  $\mathbf{R} \times ((\mathbf{R}\times \mathbf{\hat{z}})\times \mathbf{\hat{z}})$ has no component along the $z$-axis,  and the only nonzero terms in $\mathbf{F}_P$ and $\mathbf{F}_{CE}$ in section III.1 produce no torque  along the $z-$axis.  One subtle extra contribution emerges in $\mathbf{F}_P$,

\begin{eqnarray*}
  \mathbf{N}_1(P) = \int d^3 \mathbf{r}\,  \mathbf{r}\times \partial_t [\mathbf{P}(\mathbf{r},t) \times \mathbf{B}(\mathbf{r},t) ]
\end{eqnarray*}
with the polarization density given by $\mathbf{P}(\mathbf{r},t)= \mathbf{d}(t) \delta(\mathbf{r}-\mathbf{R}(t))$. The time-derivative acting on $\mathbf{d}(t)$
produces just the torque associated with the Abraham force $\mathbf{F}_A$. Without electric fields the only relevant term proportional to the velocity is
the one following from Eq.~(\ref{FVall}),
\begin{eqnarray*}
\mathbf{N}^V(1)  =  \frac{\alpha(0)}{c_0^2}  \mathbf{R} \times \partial_t ((\mathbf{V}\times \mathbf{B}_0)\times \mathbf{B}_0  )
\end{eqnarray*}
The force  proportional to $\mathbf{B}_0\times \partial_t \mathbf{B}_0$ in Eq.~(\ref{FVall}) vanishes since we have assumed $\mathbf{B}_0(t)$ to be aligned along the $z$-axis. The second contribution to $ \mathbf{N}_1(P)$ involves the time-derivative of $\mathbf{R}(t)$, and gives a torque
\begin{eqnarray*}
  \mathbf{N}^V(2) &=& \int d^3 \mathbf{r}\,  \mathbf{r}\times  [\mathbf{d}(\mathbf{r},t) \times \mathbf{B}(\mathbf{r},t) ] (-\mathbf{ V}\cdot \nabla) \delta (\mathbf{r}- \mathbf{R}(t)) \\
  &= & (\mathbf{ V}\cdot \nabla) \left\{ \mathbf{r}\times  [\mathbf{d}(\mathbf{r},t) \times \mathbf{B}(\mathbf{r},t) ]\right\} _{\mathbf{r}=\mathbf{R}(t)}
\end{eqnarray*}
with $\mathbf{V} = d\mathbf{R}/dt$. Upon inserting $\mathbf{d} = \alpha(0) (\mathbf{r} \times \partial_t\mathbf{ B}_0 )/2c_0$ we see that in this expression the gradient acts on two different position vectors $\mathbf{r}$, rather than on one which was the case when calculating the force. The one in $\mathbf{d}(\mathbf{r},t)$ has already been acknowledged in $\mathbf{N}^V(1)$, being part of the force $\mathbf{F}_V$. The extra contribution to the torque is equal to
\begin{eqnarray*}
  \mathbf{N}^V(2)  = \frac{\alpha(0)}{2c_0^2}\mathbf{V}\times  [(\mathbf{R} \times \partial_t \mathbf{B}_0)) \times \mathbf{B}_0 ]
\end{eqnarray*}
The third contribution to the torque comes from the stress tensor,
\begin{eqnarray*}
  \mathbf{N}^V(3) &= &\int d^3 \mathbf{r} \, (\mathbf{P}\times \mathbf{E} + \mathbf{M} \times \mathbf{B} ) \nonumber \\
  &=& \alpha(0) \left(\frac{\mathbf{V}}{c_0} \times \mathbf{B}_0 \right) \times \left( \frac{1}{2c_0}\mathbf{R} \times \partial_t \mathbf{B}_0 \right) \nonumber \\
  && - \left( \frac{\mathbf{V}}{c_0} \times \mathbf{d}\right) \times \mathbf{B}_0
\end{eqnarray*}
and upon inserting $\mathbf{d}= \alpha(0) \mathbf{R} \times \partial_t \mathbf{B}_0 /2c_0$ this expression can be rearranged to,
\begin{eqnarray*}
  \mathbf{N}^V(3)  = \frac{\alpha(0)}{2c_0^2} \left(
(\mathbf{ R} \times \partial_t\mathbf{B}_0) \times \mathbf{B}_0\right) \times \mathbf{V } = -\mathbf{N}^V(2)
\end{eqnarray*}
Hence,  $\mathbf{N}^V(1)$ is the only remaining external contribution from the magnetic field, together with the internal torque $\mathbf{N}_2$ in Eq.~(\ref{lenztorque}) Since its $z$-component  can be written as a full derivative involving the external, kinetic angular momentum $\mathbf{J}^{\mathrm{ext}}= \mathbf{R} \times M\mathbf{V}$ of the molecule, we conclude that the following conservation law applies,
\begin{eqnarray}\label{Nz}
 \frac{d}{dt} \left(\mathbf{J}^{\mathrm{\mathrm{ext}}} +\mathbf{J}^{\mathrm{\mathrm{int}}}  + \frac{\alpha(0)B_0^2}{Mc_0^2} \mathbf{J}^{\mathrm{ext}} - \sum_n \frac{q_n}{2c_0 }\mathbf{x}^n\times (\mathbf{x}^n \times \mathbf{B}_0) \right)_z =0 \nonumber \\
\end{eqnarray}
The conserved quantity is recognized as total  canonical angular momentum. This will become more clear in the quantum treatment discussed in the next section. The third term can be identified as an external ``Abraham angular momentum". It is small, in the former section we estimated already $\alpha(0)B^2/Mc_0^2 \approx 10^{-12} $  for a  water droplet  in 10 Tesla.

\section{Quantum treatment of polarizable bound charges}

In this section we show that the forces $\mathbf{F}_A$, $\mathbf{F}_R$ $\mathbf{F}_V$, obtained classically not without effort in previous sections and especially its magnetic components, follow from using the standard minimal coupling Hamiltonian.
In the semi-classical description of a ``molecule" containing $N$ bound charges exposed to \emph{homogeneous, time-dependent, classical } external fields $\mathbf{E}_0$ and $\mathbf{B}_0$, the  minimal-coupling Hamiltonian is given by ,
\begin{eqnarray}\label{Hminccoup}
    H =&&  \sum_{\alpha} \frac{1}{2m_\alpha}\left( \mathbf{p}_\alpha- \frac{q_\alpha}{c_0}\mathbf{A}_0(\mathbf{r}_\alpha,t)\right)^2 -  \sum_\alpha q_\alpha \mathbf{E}_0(t) \cdot \mathbf{r}_\alpha +  \nonumber \\
    && + \sum_{\alpha\neq \beta}V(|\mathbf{r}_\alpha-\mathbf{r}_\beta|)
\end{eqnarray}
in terms of the vector potential $\mathbf{A}_0(\mathbf{r}, t) =\mathbf{ B}_0(t) \times \mathbf{r}/2$ in the Coulomb gauge,  the electric field $ \mathbf{E}_0(t)$, and the Coulomb potentials $V(r)$ between the $N$ charges. In this model, the induced electric field is included since $\mathbf{E}(\mathbf{r},t) = \mathbf{E}_0(t) -\partial_t\mathbf{A}_0$, but not the induced magnetic field unless we add a term to the vector potential  $\mathbf{A}_0$ proportional to $\partial_t\mathbf{E}_0$. The operators $\mathbf{p}_{\alpha}$ and $\mathbf{r}_\alpha$ of particle $\alpha$  satisfy the canonical commutation relations $\left[{r}_{i,\alpha}, {p}_{j,\beta}\right] = i\hbar \delta_{ij} \delta_{\alpha\beta}$, even if $\mathbf{A}$ is time-dependent. The pseudo momentum $\mathbf{K}$  defined by

\begin{equation}\label{Q}
    \mathbf{K} = \sum_\alpha \mathbf{p}_\alpha + \sum_\alpha \frac{q_\alpha}{2c_0}\mathbf{B}_0 \times\mathbf{ r}_\alpha = \mathbf{P} + \frac{1}{2c_0}\mathbf{B}_0 \times\mathbf{ d}
\end{equation}
is different from\emph{ both} total canonical momentum $\mathbf{P}$ and  the total kinetic momentum
 \begin{eqnarray}\label{Qkin}
    \mathbf{P}_{kin} &=& \sum_\alpha m_\alpha \dot{\mathbf{r}}_\alpha = \mathbf{P} -
    \sum_\alpha \frac{q_\alpha}{c_0}\mathbf{A}_0(\mathbf{r}_\alpha,t)  \nonumber \\
    &=& \mathbf{P} - \frac{1}{2c_0}\mathbf{B}_0 \times\mathbf{ d}
\end{eqnarray}
 In this gauge is $ \mathbf{P}_{kin} = \mathbf{K }+  \mathbf{ d} \times \mathbf{B}_0 /c_0$. The difference between $\mathbf{P}_{kin}$ and $\mathbf{K}$ is equal to the Abraham momentum that can be identified with the force given by the integral over the molecule of Eq.~(\ref{fE}). This difference in turn was seen to be equal to the difference between the  Minkowski momentum $(4\pi c_0)^{-1}\int d^3 \mathbf{r}\,  \mathbf{D} \times \mathbf{B}$ and  the Nelson momentum $(4\pi c_0)^{-1}\int d^3   \mathbf{r}  \,\mathbf{E}\times \mathbf{B}$ \cite{AbMiBarnett}. To get the difference between Minkowski momentum and Abraham momentum $(4\pi c_0)^{-1}\int d^3  \mathbf{r}   \,\mathbf{E}\times \mathbf{H}$  in the presence of a  magnetic moment $\mathbf{m}$ of the molecule, an additional term $ c_0^{-1} \mathbf{E}_0\times\mathbf{m}  $ should be added to this difference. Indeed, Aharonov and Casher \cite{Aharonov} show from gauge invariance that $\mathbf{P}_{kin}= dL/d\mathbf{V} +\mathbf{ E}_0 \times\mathbf{ m}$ for an electrically neutral, magnetic moment moving in an electric field. This term can be included in the Hamiltonian above by adding the interaction $H_I = -\mathbf{m}\cdot (-\mathbf{P} \times \mathbf{E}_0) /Mc_0$, which is essentially the Zeeman effect in the frame co-moving with the molecule, and a force $d\mathbf{P}_{kin}/dt $ would be produced similar to Eq.~(\ref{fE2}), which would add up to the force $(2c_0)^{-1} \mathbf{m} \times \partial_t \mathbf{E}_0$ found in  Eq.~(\ref{FMQ}) from the induced magnetic field. The role of magnetic moment is here beyond the scope of the paper and will not be further addressed. It is straightforward to show that $\mathbf{K}$ commutes with $H$. To this end,
 \begin{eqnarray*}
    \frac{1}{i\hbar}\left[\mathbf{p}_{\alpha},H \right] &=& \frac{q_\alpha}{2m_\alpha c_0} \mathbf{p}_\alpha \times \mathbf{B}_0 -\frac{q^2_\alpha}{4m_\alpha}
\left( (\mathbf{B}_0\times  \mathbf{r}_\alpha) \times \mathbf{B}_0 \right) \nonumber \\
&& + q_\alpha \mathbf{E}_0 - \sum_\beta \frac{\partial V}{\partial \mathbf{r}_\alpha}
 \end{eqnarray*}
 and
  \begin{eqnarray*}
    \frac{1}{i\hbar}\left[-\frac{q_{\alpha} }{2c_0}\mathbf{B}_0 \times\mathbf{ r}_\alpha,H  \right] &=& \frac{q_\alpha}{2m_\alpha c_0} \mathbf{p}_\alpha \times \mathbf{B}_0 \nonumber \\
    &&-\frac{q^2_\alpha}{4m_\alpha}
\left( (\mathbf{B}_0\times  \mathbf{r}_\alpha) \times \mathbf{B}_0 )\right)
 \end{eqnarray*}
 The sum of both terms gives the Lorentz force $d\mathbf{p}_{kin, \alpha}/dt $ on charge $\alpha$. For the difference, summed over all charges,
 we get
   \begin{eqnarray*}
    \frac{1}{i\hbar}\left[ \mathbf{K },H \right] &=&\sum_\alpha q_\alpha \mathbf{E}_0-  \sum_{\alpha\neq \beta} \frac{\partial V(\mathbf{r}_\alpha-\mathbf{r}_\beta)}{\partial \mathbf{r}_\alpha}
    = 0
    \end{eqnarray*}
 as the total charge is zero and the two-body Coulomb interaction $V$ is symmetric in the position operators.

The time-evolution of the wave function in the presence of a time-dependent Hamiltonian is given by $
i\hbar \partial_t  |\psi(t) \rangle = H(t)  |\psi(t) \rangle$ and  $
-i\hbar \partial_t  \langle \psi(t) | =   \langle \psi(t) | H(t)$.
The time-dependence the expectation value of the time-dependent operator $\mathbf{Q}(t)$ is then
\begin{eqnarray}\label{time2}
 \nonumber   \frac{d\langle \mathbf{K}(t)\rangle}{dt} &=& \langle \psi(t) |  \left( \frac{1}{i\hbar}[\mathbf{Q},H] + \frac{\partial \mathbf{K}}{\partial t} \right) | \psi(t) \rangle \\
&=& \langle \psi(t) | \left( \frac{1}{i\hbar}[\mathbf{K},H] + \frac{\partial \mathbf{K}}{\partial t} \right) | \psi(t) \rangle
\end{eqnarray}
 Since $[\mathbf{K},H] =0 $ at all times, it follows that the pseudo-momentum $\mathbf{K}$ is  conserved  in a time-independent magnetic field. In general,
 \begin{equation}\label{Qt}
    \frac{d\langle \mathbf{K}(t)\rangle}{dt}  =   \frac{1}{2c_0}  \frac{d\mathbf{B}_0}{dt} \times \langle\mathbf{d }\rangle
\end{equation}
  Recalling the definition of $\mathbf{K}$ in terms of kinetic momentum this implies
  \begin{eqnarray}\label{KINt}
  \nonumber \frac{d\langle\mathbf{P}_{kin}\rangle}{dt} &=&    \frac{d}{dt} \left( \, \langle\mathbf{d }\rangle  \times \mathbf{B}_0 \right)  - \frac{1}{2c_0}\langle\mathbf{d } \rangle \times \frac{d\mathbf{B}_0}{dt}\\
  &=&   \frac{1}{c_0}\frac{d\langle\mathbf{d }\rangle}{dt}  \times \mathbf{B}_0  + \frac{1}{2c_0}\langle\mathbf{d } \rangle \times \frac{d\mathbf{B}_0}{dt}
\end{eqnarray}
This equation is generally valid as long as the fields vary slowly. In the presence of an electric field, first-order perturbation theory with the interaction  $-q_\alpha \mathbf{r}_\alpha \cdot \mathbf{E}_0$ in the Hamiltonian~(\ref{Hminccoup}) leads to $\left<\mathbf{d} (t)\right> = \alpha(0) \mathbf{E}_0(t)$ for slowly varying fields, and we recover the classical Abraham force obtained in Eq.~(\ref{Fa1}), with asymmetrical roles for $\mathbf{E}_0$ and $\mathbf{B}_0$.

Equation~(\ref{KINt}) is gauge-invariant.  To obtain the Abraham force on a moving, neutral molecule subject to a time-dependent magnetic field only, it is instructive to perform the gauge-transformation $\mathbf{A}(\mathbf{r},t) \rightarrow \mathbf{A}(\mathbf{r},t) + \nabla\chi(\mathbf{r},t) $ and $\phi(\mathbf{r},t) \rightarrow \phi(\mathbf{r},t) - c_0^{-1}\partial_t \chi(\mathbf{r},t)$, where we choose $\chi(\mathbf{r},t) =\mathbf{ r}\cdot (\mathbf{R} \times \mathbf{B}_0(t))/2$ with $\mathbf{R}$ the center-of-mass position operator of the molecule \cite{cohengaugeR}.  If we use that $\sum_\alpha q_\alpha =0$, the Hamiltonian transforms to,
\begin{eqnarray}\label{Hdipolar}
    \tilde{H} =&&  \sum_{\alpha} \frac{1}{2m_\alpha}\left( \frac{m_\alpha}{M} \mathbf{P }+ \mathbf{\delta p}_\alpha \right. \nonumber \\ && \left. - \frac{q_\alpha}{2c_0} (\mathbf{B}_0(t) \times \delta \mathbf{r}_\alpha) + \frac{m_\alpha}{2Mc_0} \mathbf{d}\times \mathbf{B}_0(t) \right)^2  \nonumber \\ -&&    \frac{1}{2c_0}\mathbf{d} \cdot \left[\mathbf{R} \times \partial_t \mathbf{B}_0(t) \right]
     + \sum_{\alpha\neq \beta}V(|\delta \mathbf{r}_\alpha-\delta \mathbf{r}_\beta|)
\end{eqnarray}
We introduced the internal distances $ \delta\mathbf{ r}_\alpha = \mathbf{r}_\alpha - \mathbf{R}$ of charge $\alpha$ to the center of mass, and separated  canonical momentum of the center-of-mass motion $\mathbf{P}= \sum_\alpha  \mathbf{p}_\alpha$ from the internal canonical momenta $\delta \mathbf{p}_\alpha$ according to $\mathbf{p}_\alpha = m_\alpha \mathbf{P}/M  + \delta \mathbf{p}_\alpha$. Note that the dipole operator $\mathbf{d}= \sum_\alpha q_\alpha \mathbf{r}_\alpha$  does not depend on the location $\mathbf{R}$ of the center of mass, since the molecule is globally neutral. The second line in the equation above
would be absent in the widely applied long- wavelength approximation $\mathbf{A}_0(\mathbf{r}_\alpha,t) \approx \mathbf{A}_0(\mathbf{R},t)$ and is here linear in $\delta \mathbf{r}_\alpha$.
 It is well-known that the  $N$ internal momenta $\delta \mathbf{p}_\alpha$ and $N$ internal coordinates $\delta \mathbf{r}_\alpha$ can be written in terms of $N-1$ independent momenta and position vectors that satisfy the canonical commutation relations.  In the new gauge we identify the total kinetic momentum as $\mathbf{P}_{kin} = M[\mathbf{R},\tilde{H}]/i\hbar = \mathbf{P} - \mathbf{B}_0 \times \mathbf{d}/c_0$, independent of $\mathbf{R}$, as also found in Ref.~\cite{loudonradiative}. Since $[\mathbf{P}, \tilde{H}] =\mathbf{ d} \times \partial_t\mathbf{ B}_0/2c_0 $ we recover the gauge-invariant expression~(\ref{KINt}).
 Since $\sum_\alpha \delta \mathbf{p}_\alpha =0$, we can write $\tilde{H}= \tilde{H}_{at} + \mathbf{P}^2/2M + \tilde{H}_{I}$, with the interaction between internal and external motion in the new gauge is given by,
\begin{eqnarray}\label{HintV}
    \tilde{H}_I &= &  -\mathbf{d} \cdot \left(  \frac{1}{Mc_0}  \mathbf{P } \times \mathbf{B}_0   +   \frac{1}{2c_0}\   \mathbf{R} \times \partial_t \mathbf{B}_0(t) \right) \nonumber
\end{eqnarray}
 To calculate the expectation value of the electric dipole operator $\mathbf{d}$ subject to the time-dependent interaction $\tilde{H}_I(t) \equiv -\mathbf{d}\cdot \mathbf{O}(t)$, a first and good approximation is to decouple the expectation values of internal ($\mathbf{d}$) from external operators ($\mathbf{R}$ and $\mathbf{P}$). This is justified when the internal atomic energy level spacings $E_n-E_0$ of $\tilde{H}_{at}$ are hardly affected by the external motion described by $\mathbf{P}^2/2M$ . Similarly, typical frequencies involved in the time-dependence of $\mathbf{B}_0$ are supposed to obey $\hbar \omega \ll (E_n-E_0)$.
 Under these circumstances, we get $ \langle \mathbf{d }(t)\rangle = \alpha(0) \cdot \langle \mathbf{O}(t) \rangle$.  Applying this to Eq.~(\ref{KINt}) gives
\begin{eqnarray}\label{FVquantum}
\frac{d\langle\mathbf{P}_{kin}\rangle}{dt} = \mathbf{F}_V+ \mathbf{F}_R
\end{eqnarray}
with
\begin{eqnarray}
\mathbf{F}_V&=& \frac{\alpha(0)}{Mc_0^2}\frac{d}{dt} \left( (\langle\mathbf{P }\rangle\times \mathbf{B}_0) \times \mathbf{B}_0 \right) \nonumber \\
&& \ \ - \frac{\alpha(0)}{2Mc_0^2} (\mathbf{P }\times \mathbf{B}_0) \times \partial_t\mathbf{B}_0 \nonumber \\
&& \ \ + \frac{\alpha(0)}{2c_0^2} (\mathbf{V} \times \partial_t\mathbf{B}_0) \times \mathbf{B}_0 \nonumber \\
&=& \frac{\alpha(0)}{Mc_0^2}\frac{d}{dt} \left( (\langle\mathbf{P }\rangle\times \mathbf{B}_0) \times \mathbf{B}_0 \right)
\nonumber \\  && \ \  + \frac{\alpha(0)}{2c_0^2} \mathbf{V} \times (\partial_t\mathbf{B}_0 \times \mathbf{B}_0 )
\end{eqnarray}
where $\mathbf{V}= d \langle \mathbf{R} \rangle /dt$. Approximating $\langle \mathbf{P }\rangle = \langle\mathbf{P}_{kin}\rangle + \mathcal{O}(\mathbf{B}_0) = M\mathbf{V}$, makes this outcome coincide with the classical derivation that had led us to Eq.~(\ref{FVall}). The force $\mathbf{F}_R$ is,
\begin{eqnarray}
\mathbf{F}_R&=& \frac{\alpha(0)}{2c_0^2}   (\langle\mathbf{R }\rangle\times \partial_t^2 \mathbf{B}_0) \times \mathbf{B}_0   \nonumber \\
&& \ \ + \frac{\alpha(0)}{4c_0^2} (\mathbf{R}\times \partial_t\mathbf{B}_0) \times \partial_t\mathbf{B}_0
\end{eqnarray}
and equal to Eq.~(\ref{FRmagnetic}).

\section{Quantum torque on chiral molecule}

In this section we derive, using the minimal coupling Hamiltonian, the Abraham-type torque on a chiral particle, found classically in Eq.~(\ref{chiral}) in the presence of time-dependent electromagnetic fields.
The Hamiltonian is
\begin{eqnarray}\label{Hminccoup2}
    H =&&  \sum_{\alpha} \frac{1}{2m_\alpha}\left( \mathbf{p}_\alpha- \frac{q_\alpha}{c_0}\mathbf{A}_0(\mathbf{r}_\alpha,t)\right)^2  - \mathbf{d}\cdot \mathbf{E}_0(t) +  \nonumber \\
    && +  \sum_{\alpha\neq \beta}V(|\mathbf{r}_\alpha-\mathbf{r}_\beta|)
\end{eqnarray}
with $\mathbf{A}(\mathbf{r},t) = \mathbf{B}_0(t) \times \mathbf{r}/2$.  It follows straightforwardly that the total canonical momentum  $\mathbf{J}= \sum_\alpha \mathbf{r}_\alpha \times \mathbf{p}_\alpha$ obeys
\begin{equation}\label{AMconser}
    \frac{d\mathbf{J}}{dt} =  \frac{1}{i\hbar} [\mathbf{J}, H] = \mathbf{m} \times\mathbf{ B} + \mathbf{d }\times \mathbf{E }+ \sum_\alpha \frac{q_\alpha^2}{4 m_\alpha} (\mathbf{B}_0\cdot \mathbf{r}_\alpha) \mathbf{r}_\alpha \times\mathbf{ B}_0
\end{equation}
with  electric dipole moment operator $\mathbf{d}=\sum_\alpha q_\alpha \mathbf{r}_\alpha$ and magnetic dipole moment operator
$\mathbf{m}=  \sum_\alpha (q_\alpha/2m_\alpha)  \mathbf{r}_\alpha \times \mathbf{p}_\alpha $.  The third term above vanishes for rotational symmetry and will be ignored. We calculate the quantum expectation value of $\mathbf{J}$ with the time-dependent interaction.
\begin{equation}\label{intH3}
    H_I(t) = -\mathbf{d}\cdot \mathbf{E}_0(t) - \mathbf{m}\cdot \mathbf{B}_0(t)
\end{equation}
Clearly the usual relations $\left\langle \mathbf{d}\right\rangle = \alpha(0) \mathbf{E}_0(t)$ and $\left\langle \mathbf{m}\right\rangle = \chi(0) \mathbf{B}_0(t)$ give no contributions to the torque in Eq.~(\ref{AMconser}). If we switch on the magnetic perturbation slowly at $t=0$ for an molecule being in eigenstate $a$, first-order time-dependent perturbation theory gives for $\left\langle \mathbf{d}\right\rangle (t)$,
\begin{eqnarray*}
&& \left\langle \mathbf{d}\right\rangle(t)  = \sum_{n\neq a} \langle a | \mathbf{d} |n\rangle \exp\left(-\frac{i}{\hbar}(E_n-E_a)t\right)  \\ && \times
  \int_0^t dt'
\exp\left(\frac{i}{\hbar}(E_n-E_a) t'\right)) \langle n | -\mathbf{m} | a \rangle \cdot \mathbf{B}_0(t')  + c.c   \\
\end{eqnarray*}
Upon performing one integration by parts,
\begin{eqnarray*}
&& \left\langle \mathbf{d}\right\rangle(t) = \sum_{n\neq a} \frac{\langle a | \mathbf{d} |n\rangle }{E_a-E_n} \langle n |\mathbf{m} | a \rangle \cdot \mathbf{B}_0(t) \\ &+& \sum_{n\neq a} \frac{\langle a | \mathbf{d} |n\rangle }{E_a-E_n}  \exp\left(-\frac{i}{\hbar}(E_n-E_a) t'\right)  \times \\  && \int_0^t dt'
\exp\left(\frac{i}{\hbar}(E_n-E_a) t'\right)  \langle n |\mathbf{m} | a \rangle \cdot \partial_t \mathbf{B}_0(t')
   + c.c
\end{eqnarray*}
In the absence of a magnetic field (which would here be a higher order effect), $\langle a | \mathbf{d} |n\rangle$ is real-valued whereas $\langle a | \mathbf{m} |n\rangle$ is purely imaginary. The first term, that would stand for the adiabatic perturbation of the dipole moment by the magnetic field, is thus canceled by its complex conjugation. One more integration by parts leads to
\begin{eqnarray*}
\left\langle \mathbf{d}\right\rangle(t) = \frac{1}{c_0}\mathbf{g}(0)
 \cdot \partial_t \mathbf{B}_0
\end{eqnarray*}
with the static rotatory power defined by \cite{Craig}
\begin{equation}\label{RPDEF}
    \mathbf{g}(0)= \frac{\hbar c_0}{i}\sum_{n\neq a} \frac{\langle a | \mathbf{d} |n\rangle \langle n |\mathbf{m} | a \rangle}{(E_a-E_n)^2}  + c.c
\end{equation}
and where higher time-derivatives of $\mathbf{B}_0$ have been neglected. The magnetization induced by the electric field involves the same calculation,
but now features $  \langle a | \mathbf{m} |n\rangle \langle n |\mathbf{d} | a \rangle = - \langle a | \mathbf{d} |n\rangle \langle n |\mathbf{m} | a \rangle$. Hence
\begin{eqnarray*}
\left\langle \mathbf{m}\right\rangle(t) = -\frac{1}{c_0}\mathbf{g}(0)
 \cdot \partial_t \mathbf{E}_0
\end{eqnarray*}
In the presence of rotational symmetry, $g(0)$ reduces to a scalar. From  Eq.~(\ref{AMconser})   we find
\begin{equation}\label{RPtorquequantum}
    \frac{d\mathbf{J}}{dt} = \frac{g(0)}{c_0} \frac{d}{dt}  (\mathbf{E}_0 \times \mathbf{B}_0)
\end{equation}
Since the difference between canonical and kinetic momentum is directed along the $\mathbf{B}_0$-axis, we conclude that the kinetic momentum along the
$s$-axis, defined by the vector $\mathbf{S}=\mathbf{E}_0 \times \mathbf{B}_0$,  obeys the conservation law $J^{kin}_s - g(0)E_0(t)B_0(t)/c_0=$ constant.

\section{Quantum torque on moving polarizable molecule}

In this section we derive quantum-mechanically the conservation of canonical angular momentum, expressed by Eq.~(\ref{Nz}), of a neutral molecule in a
homogeneous, slowly varying magnetic field. Let us consider $N$ charges circulating in an external central field $W(R)$.
The Hamiltonian is
\begin{eqnarray}\label{Hminccoup3}
    H =&&  \sum_{\alpha} \frac{1}{2m_\alpha}\left( \mathbf{p}_\alpha- \frac{q_\alpha}{c_0}\mathbf{A}_0(\mathbf{r}_\alpha,t)\right)^2  +  \nonumber \\
    && + \sum_{\alpha\neq \beta}V(|\mathbf{r}_\alpha-\mathbf{r}_\beta|) + W(R)
\end{eqnarray}
with $\mathbf{R}$ the center of mass and again $\mathbf{A}(\mathbf{r},t) = \mathbf{B}_0(t) \times \mathbf{r}/2$. We assume that the molecule rotates as a whole in the potential field, characterized by an external angular momentum that is not necessarily conserved. The total canonical angular momentum is given by $\mathbf{J}= \sum_\alpha \mathbf{r}_\alpha \times \mathbf{p}_\alpha$. We choose $\mathbf{B}_0(t)$ to be constantly aligned along the $z$-axis.
It can be verified that $[J_z, H] =0$, and since there is no explicit time-dependence in $J_z$ we immediately conclude that total canonical momentum is conserved,
\begin{eqnarray}\label{Jkinconserved1}
    \frac{d}{dt} \left(\mathbf{J}_{kin} + \sum_\alpha \frac{q_\alpha}{2c_0} \mathbf{r}_\alpha \times (\mathbf{B}_0(t) \times \mathbf{r}_\alpha   ) \right)_z =0
\end{eqnarray}
This shows that $J_z$ must be gauge-invariant up to a time-independent term.
Inserting $\mathbf{r}_\alpha = \mathbf{R} + \delta \mathbf{r}_\alpha$, with $\mathbf{R}$ the center-of-mass operator, gives
 \begin{eqnarray}\label{Jkinconserved2}
    \frac{d}{dt} \left(J_{kin, z} + \frac{B_0}{c_0} (\mathbf{\mathbf{R}}_\parallel \cdot \mathbf{d}_\parallel)
   +\frac{B_0(t)}{2c_0} \sum_\alpha  q_\alpha
    \delta \mathbf{r}_{\alpha,\parallel}^2   \right) =0\nonumber \\
\end{eqnarray}
with  operators  $\mathbf{\mathbf{R}}_\parallel$, $\delta \mathbf{r}_{\alpha,\parallel}$ and electric-dipole operator $ \mathbf{d}_\parallel$  projected in the plane perpendicular to $\mathbf{B}_0$.

We will show the equivalence to Eq.~(\ref{Nz}) and also establish that internal and external canonical angular momentum are conserved separately in this simple model. To this end it is again convenient to use the new gauge with Hamiltonian~(\ref{Hdipolar}), using the unitary transformation
$T(t) = \exp(-i\mathbf{d}\cdot \mathbf{A}(\mathbf{R},t)/\hbar ) $. In the new gauge canonical and kinetic momentum of particle $\alpha$ are connected by
\begin{equation}\nonumber
   \mathbf{ p}_\alpha^{kin} = \mathbf{p}_\alpha + \frac{q_\alpha}{2c_0} \delta \mathbf{r}_\alpha \times \mathbf{B}_0 + \frac{m_\alpha}{2Mc_0} \mathbf{d}\times \mathbf{B}_0
\end{equation}
It is indeed verified that $T(t){J}_zT(t)^* =  {J }_z$.
It also follows  that
\begin{equation}\label{Jkincan}
    \mathbf{J}_{kin} = \mathbf{J} + \frac{1}{c_0} \left[\mathbf{R} \times (\mathbf{d}\times \mathbf{B}_0) \right] + \sum_\alpha  \frac{q_\alpha }{2c_0}
    \delta \mathbf{r}_\alpha \times  (\delta \mathbf{r}_\alpha \times \mathbf{B}_0)
\end{equation}

We have from Eq.~(\ref{Hdipolar}) $\tilde{H} = T(t)HT(t)^*  + \mathbf{d} \cdot \partial_t \mathbf{A}(\mathbf{R},t)$
From this
\begin{eqnarray*}
   [\tilde{J}_z,\tilde{H}] &=& [\tilde{J}_z,T(t){H}T^(t)^* + \mathbf{d} \cdot \partial_t \mathbf{A}(\mathbf{R},t)]  \\
  & =& T(t)[{J}_z,{H}] T(t)^* + [ {J}_z, \mathbf{d} \cdot \partial_t \mathbf{A}(\mathbf{R},t)]  \\
  & = & [ {J}_z, \mathbf{d} \cdot \partial_t \mathbf{A}(\mathbf{R},t)] =  0
\end{eqnarray*}
The final equality follows straightforwardly by inserting $\mathbf{A} = \sum_\beta (m_\beta/2M) \mathbf{B}_0(t) \times \mathbf{r}_\beta$. Using Eq.~(\ref{Jkincan}) and $[\tilde{J}_z,\tilde{H}] =0$  we recover the conservation of total canonical angular momentum expressed by Eq.~(\ref{Jkinconserved2}). We can look separately at the external angular momentum $L_z =(\mathbf{R} \times \mathbf{P})_z$,
\begin{eqnarray*}
  \frac{1}{i\hbar} [L_z,\tilde{H}_I] =  \frac{B_0 }{Mc_0} (\mathbf{P}_\parallel\cdot \mathbf{d}_\parallel) +  \frac{\partial_t B_0 }{2c_0} (\mathbf{R}_\parallel\cdot \mathbf{d}_\parallel)
\end{eqnarray*}
We can calculate the quantum expectation value of the operators $(\mathbf{P}_\parallel\cdot \mathbf{d}_\parallel)$ and $(\mathbf{R}_\parallel\cdot \mathbf{d}_\parallel)$. To this end we write $\tilde{H}(t) = H_0 + H_e+ \tilde{H}_I(t)$, with $H_e= \mathbf{P}^2/2M + W(R)$, and assume at $t=0$ that $\tilde{H}_I(t=0)$ and the molecule to be in the eigenstate  $|a\rangle \otimes |L_z\rangle$. Since the time-dependence is slow (compared to the level spacing) the molecule follows the  time-dependent interaction~(\ref{HintV}) adiabatically, and first-order perturbation theory gives,
\begin{eqnarray*}
  &&\left\langle (\mathbf{R}_\parallel \cdot \mathbf{d}_\parallel) \right \rangle  = \\
 && \langle a,L_z |  (\mathbf{R}_\parallel \cdot \mathbf{d}_\parallel) \frac{1}{ {E}_a +E_{L_z} -\tilde{H}_0 -H_e} \tilde{H}_I(t) | a, L_z  \rangle  + c.c  \\
\end{eqnarray*}
It is reasonable to assume that the  external motion $H_e$ perturbs only weakly the energy levels of $H_0$, since $W(R)$ is small compared to the Coulomb interactions $V(r)$. In that case internal and external quantum expectations decouple and
\begin{eqnarray*}
 &&  \left\langle  (\mathbf{R}_\parallel \cdot \mathbf{d}_\parallel) \right\rangle  \\
 &= & \epsilon_{klz} \frac{B_0}{Mc_0}  \langle  L_z|  R_i P_l | L_z \rangle \sum_{n\neq a}  \frac{\langle 0|d_i | n \rangle \langle n| d_k |0  \rangle}{ {E}_n - {E}_a}    + c.c  \\
 &=& \frac{B_0}{Mc_0} \epsilon_{klz}  \langle  L_z|  R_i P_l | L_z \rangle  \alpha_{ik}(0)
=  \frac{\alpha(0) B_0}{Mc_0} L_z
\end{eqnarray*}
The last equality applies for an isotropic electrical polarizability.
Similarly,
\begin{eqnarray*}
  \left\langle  (\mathbf{P}_\parallel \cdot \mathbf{d}_\parallel) \right\rangle =  -\frac{\alpha(0) \partial_t B_0}{2c_0} L_z
\end{eqnarray*}
Inserting this, we find that the expectations of the two term of the commutator $[L_z,\tilde{H}_I] $ cancel and
\begin{eqnarray*}
  \frac{d}{dt}\left\langle  L_z \right\rangle = \frac{d}{dt}\left(\langle   L_z^{kin} \rangle +  \frac{\alpha(0) B^2_0(t)}{Mc_0^2} L_z  \right) =0
\end{eqnarray*}
valid up to factors $B_0^2(t)$. We see that in first-order perturbation theory Eq.(\ref{Jkinconserved2}) becomes equal to Eq.~(\ref{Nz}), and that Eq.(\ref{Jkinconserved2}) splits up into a separate conservation law for external and internal canonical angular momentum. The total canonical angular momentum is rigorously conserved in all orders of perturbation theory.

\section{Conclusions}

In this work we have discussed generalized Abraham forces and torques induced by slowly varying electromagnetic fields on polarizable matter. We have provided both a classical and a quantum-mechanical derivation. These forces are small but touch essential features of electromagnetism. We have found that for a molecule, due to induced fields,  the ``standard Abraham force" , $\mathbf{F}_A = c_0^{-1}d/dt  \, (\mathbf{\mathbf{d}}  \times \mathbf{\mathbf{B}})$  in terms of electric dipole moment and magnetic field splits up into unequal time-derivatives of magnetic and electric fields. In the quantum formulation, this asymmetry arises because pseudo-momentum is no longer conserved in a time-dependent magnetic field.  For a molecule moving at speed $\mathbf{V}$, a force $\mathbf{F}_V = \alpha(0) d/dt [(\mathbf{V}\times \mathbf{B})\times \mathbf{B}]/c_0 $ exists as a full derivative with time, with $\alpha(0)$ the static polarizability, but without an electric equivalent. This asymmetry is even more remarkable and calls for the existence of a ``dual" force  $\mathbf{F}_M = c_0^{-1}d/dt \, (\mathbf{\mathbf{E}}  \times \mathbf{\mathbf{m}})$, with $\mathbf{m}$ the magnetic moment. The purpose of this work was not to develop such dual theory, but rather to argue the need for it.  Propositions for dual extensions of Maxwell equations exist already \cite{dual,dual2}, and indeed,this dual force has been already suggested in literature \cite{AbMiBarnett}, here confirmed, inspired by the fundamental link with the Aharonov-Casher topological phase \cite{Aharonov} of a neutral particle with a magnetic moment. We have also found that rotatory power of a molecule, induced by its broken mirror symmetry, generates an Abraham torque equal to $g(0) d/dt( \mathbf{E} \times \mathbf{B})/c_0$, with $g(0)$ the static rotatory power. In this case the symmetry between magnetic and electric fields is explicit in the constitutive equations $\mathbf{d}= g(0) \partial_t \mathbf{B}$ and $\mathbf{m} = -g(0) \partial_t \mathbf{E}$. Finally, a neutral molecule with a finite angular momentum $J_z$ and subject to a time-dependent homogeneous magnetic field in the same direction, experiences an Abraham torque equal to $\alpha(0)d/dt(B^2J_z)/Mc_0^2$. In both cases the Abraham torque can be understood from a generalized canonical momentum, which is quite clear in the quantum-mechanical formulation.

The author thanks Geert Rikken for useful discussions.  

\end{document}